\documentclass[twocolumn,pra,superscriptaddress,noshowpacs,epsf]{revtex4}

\usepackage{graphicx}

\usepackage{epsfig}
\usepackage[10pt]{moresize}
\usepackage{amssymb}
\usepackage{comment}
\usepackage{amsmath}

\usepackage{amscd,epsfig,graphicx,rotate}
\usepackage{color}
\usepackage{epsf}
\usepackage[english]{babel}
\usepackage{mathrsfs}
\usepackage[hypertex]{hyperref}
\usepackage{pifont,bm}
\usepackage{amsfonts}

\begin{document}

\title{Encoding a qubit with Majorana modes in superconducting circuits}
\date{\today}
\author{J. Q. You}
\affiliation{Beijing Computational Science Research Center, Beijing
100084, China}
\affiliation{Center for Emergent Matter Science, RIKEN, Wako-shi 351-0198, Japan}

\author{Z. D. Wang}
\affiliation{Department of Physics and Center of Theoretical and
Computational Physics, The University of Hong Kong, Pokfulam Road,
Hong Kong, China}

\author{Wenxian Zhang}
\affiliation{School of Physics and Technology, Wuhan University, Wuhan 430072, China}

\author{Franco Nori}
\affiliation{Center for Emergent Matter Science, RIKEN, Wako-shi 351-0198, Japan}
\affiliation{Physics Department, The University of Michigan, Ann Arbor,
MI 48109-1040, USA}

\begin{abstract}
Majorana fermions are long-sought exotic particles that are their
own antiparticles. Here we {propose to utilize} superconducting circuits to construct two
superconducting-qubit arrays where Majorana modes can occur. {A
so-called Majorana qubit is encoded by using the unpaired Majorana modes, which emerge at the
left and right ends of the chain in the Majorana-fermion representation. We also show this Majorana qubit in the spin representation and its advantage, over a single superconducting qubit, regarding quantum coherence.} Moreover, we propose to use four superconducting
qubits as the smallest system to demonstrate the braiding of
Majorana modes and show how the states before and after braiding
Majoranas can be discriminated.
\end{abstract}

\maketitle

Majorana fermions are particles that are their own antiparticles.
These long-sought particles have recently received considerable
interest (see, e.g., Refs.~\cite{Wilczek,Stern,Read,Ivanov,Kitaev-1,Fu,Sato,Sau,Alicea-PRB,Lee,Rakhmanov}).
It has been recognized~\cite{wire1,wire2,wire3} that a relatively
easy-to-engineer system{\bf---}one-dimensional (1D) semiconducting
wires on an $s$-wave superconductor{\bf---}can realize a nontrivial
topological state supporting Majorana fermions. This state is
characteristic of 1D topological superconductors~\cite{Kitaev-1}, in
which Majorana modes can occur without requiring the presence of
vortices in the system. {The recent experimental observation~\cite{Mourik} of a zero-bias peak
in the differential conductance of a semiconductor nanowire coupled to a superconductor
suggested the possible existence
of Majorana fermions.} Moreover, it was proposed~\cite{Alicea} to
use tunable 1D semiconducting wire networks on an $s$-wave
superconductor to demonstrate the non-Abelian statistics of Majorana
fermions, because the Majoranas in the semiconducting wires can also
behave like vortices in a $p+ip$
superconductor{~\cite{Read,Ivanov}}. In addition, it was also
recognized~\cite{Kitaev-2, Lieb} that when a Jordan-Wigner
transformation is performed, a 1D quantum Ising model is equivalent
to a 1D topological superconductor, and Majorana modes can also
occur therein. Nevertheless, less attention has been paid to this
quantum Ising model than to 1D topological superconductors because
it was often regarded as a toy model.

In this paper, we {propose to realize such a toy model by using experimentally accessible superconducting-qubit arrays.}
Importantly, superconducting qubits can behave as controllable
artificial atoms and tunable interqubit couplings are also
achievable (see, e.g., Refs.~\cite{You-Nature,You-PT,Clarke}).
{For instance, the tunable coupling between flux qubits was experimentally demonstrated
in Refs.~\cite{D-wave,Hime, Tsai}.} For a finite
superconducting-qubit array, when the interqubit couplings are tuned
to be nonzero and other parameters of the qubits are tuned to be
zero, there are two unpaired Majorana modes, {which emerge at the left and right
ends of the chain in the Majorana-fermion representation. We use these two Majorana modes to encode a
qubit which is here called the Majorana qubit. Also, we express this Majorana qubit in the spin representation and show its advantage, over a single superconducting qubit, regarding quantum coherence.} Moreover,
the advantages of superconducting qubits in controllability make it
possible to construct a tunable 1D quantum Ising model on wire
networks, similar to the semiconducting wire networks in
Ref.~\cite{Alicea}, to demonstrate the non-Abelian statistics of Majorana
modes. We propose to use
four superconducting qubits as the smallest circuit to demonstrate
the braiding of Majorana modes, and show how the states before and
after braiding Majoranas can be discriminated. This should provide
an experimentally realizable, relatively simple setup to manipulate
and probe Majorana fermions.
Thus, our proposal could allow the quantum simulation~\cite{Buluta}
or emulation of Majorana fermions.

\vspace{.8cm}
\noindent{\large\bf Results}

\vspace{.1cm}
\noindent
{\bf Majorana modes in superconducting circuits.~}We construct two types of superconducting-qubit arrays (see
Figure~1), which can exhibit Majorana modes.

\vspace{.2cm}
\noindent
{\it (1)~Charge-qubit array.} For the array of charge qubits shown in Figure~1(a), every
pair of nearest-neighbor qubits are coupled by a large Josephson
junction acting as an effective inductance. The non-nearest-neighbor
qubits can also be coupled via these large Josephson junctions, but
the interactions are negligibly small. Here we assume that all
charge qubits are identical and that all large junctions are equal
to each other.
When leading terms are considered, the Hamiltonian of this
charge-qubit array can be written as
\begin{equation}
H=\sum_{n=1}^{N-1} t\,\sigma_n^x\sigma_{n+1}^x -\sum_{n=1}^N
\left(\mu\,\sigma_n^z+\nu\,\sigma_n^x\right), \label{H-1}
\end{equation}
with $\mu=\frac{1}{2}E_{\rm ch}(1-C_gV_g/e)$,
$\nu=E_{J0}\cos(\pi\Phi_q/\Phi_0)$, and the interqubit
coupling is given by~\cite{YTN-PRB}
\begin{equation}
t=L_J\left(\frac{\pi E_{J0}}{\Phi_0}\right)^2
\sin^2\left(\frac{\pi\Phi_q}{\Phi_0}\right).
\label{T-1}
\end{equation}
Here $E_{\rm ch}$ ($\approx e^2/C_J$) $\gg E_J$ in the charging
regime considered here and $L_J={\Phi_0}/{2\pi I_c}$, with $I_c=2\pi
E_{Jc}/\Phi_0$ and $\Phi_0$ being the flux quantum. The eigenstates
of the Pauli operator $\sigma_n^z$ are the charge states
$|0_n\rangle$ and $|1_n\rangle$, corresponding to zero and one extra
Cooper pair in the superconducting island of the $n$th qubit.
The Hamiltonian (\ref{H-1}) provides an analog to the 1D quantum
Ising model.

\begin{figure}
\includegraphics[width=3.3in,
bbllx=36,bblly=150,bburx=556,bbury=721]
{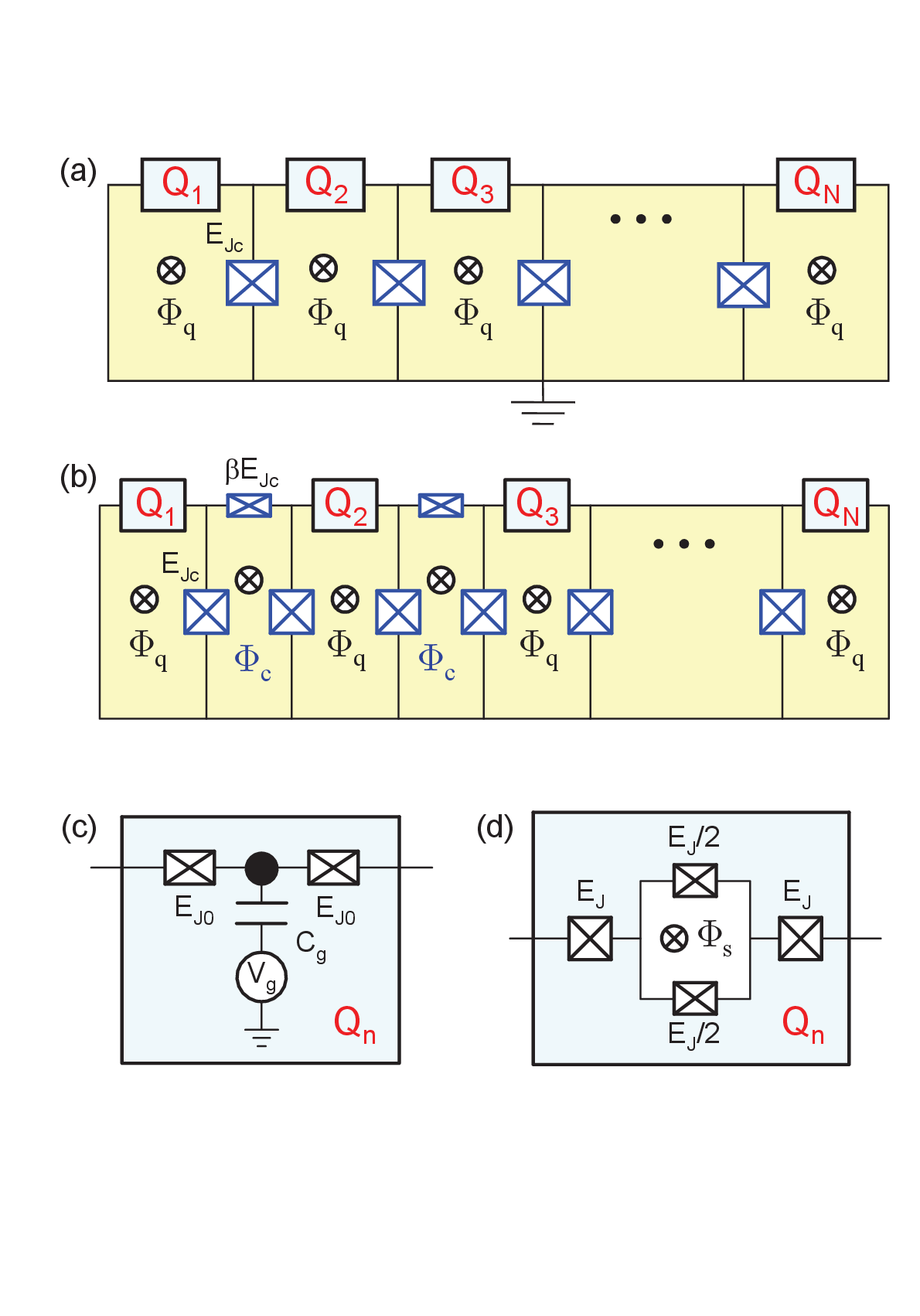}\caption{{\bf Two arrays of superconducting qubits.}
(a)~Charge-qubit array: Nearest-neighbor charge qubits $Q_n$ and $Q_{n+1}$ are
coupled by a large Josephson junction with coupling energy $E_{Jc}$
(shown as a crossed rectangle). (b)~Flux-qubit array:
Nearest-neighbor flux qubits are coupled by a coupler consisting of
a flux-biased loop that is interrupted by two large Josephson
junctions (each with coupling energy $E_{Jc}$) and a small Josephson
junction with coupling energy $\beta E_{Jc}$, where $0<\beta\ll 1$.
In (a) and (b), $\Phi_q$ is the flux applied to each qubit loop.
(c)~Main components of a charge qubit, where a superconducting
island (denoted as a solid circle) is connected to two Josephson
junctions (each with coupling energy $E_{J0}\ll E_{Jc}$ and
capacitance $C_J$) and biased by a voltage $V_g$ through a gate
capacitance $C_g\ll C_J$. (d)~Main components of a flux qubit, where
two Josephson junctions, each with coupling energy $E_J\ll E_{Jc}$,
connects a symmetric dc SQUID biased by a flux $\Phi_s$.
} \label{fig1}
\end{figure}

We now consider the case with the fluxes in all
charge-qubit loops being tuned to $\Phi_q=\frac{1}{2}\Phi_0$, so
that $\nu=0$, and the interqubit couplings reach the maximum
$t=L_J(\pi E_{J0}/\Phi_0)^2$. Using the Jordan-Wigner
transformation~\cite{Kitaev-2,Lieb}:
\begin{equation}
a_n=\sigma_n^-\prod_{m=1}^{n-1}\sigma_m^z,~~~~
a_n^{\dag}=\sigma_n^+\prod_{m=1}^{n-1}\sigma_m^z,
\end{equation}
where $\sigma_n^{\pm}=\frac{1}{2}(\sigma_n^x\pm i\sigma_n^y)$, one
can cast equation~(\ref{H-1}), in the case of
$\Phi_q=\frac{1}{2}\Phi_0$, to
\begin{equation}
H=\sum_{n=1}^{N-1}t(a_n-a_n^{\dag})(a_{n+1}+a_{n+1}^{\dag})
-\sum_{n=1}^N\mu(2a_n^{\dag}a_n-1),
\label{HF}
\end{equation}
where the Dirac fermions obey the anticommutation relation
$\{a_n,a_{n'}^{\dag}\}=\delta_{nn'}$. Introducing Majorana fermions:
\begin{equation}
\gamma_n^A=a_n^{\dag}+a_n, ~~~\gamma_n^B=i(a_n^{\dag}-a_n),
\label{MF}
\end{equation}
one can rewrite the Hamiltonian (\ref{HF}) as
\begin{equation}
H=i\sum_{n=1}^{N-1}t\gamma_n^B\gamma_{n+1}^A-i\sum_{n=1}^N\mu
\gamma_n^A\gamma_n^B, \label{HMF}
\end{equation}
where $\gamma_n^{X\dag}=\gamma_n^X$ and
$\{\gamma_n^X,\gamma_{n'}^{X'}\}=2\delta_{XX'}\delta_{nn'}$.
Obviously, $(\gamma_n^X)^2=1$, which is different from the Dirac
fermion.

\vspace{.2cm}
\noindent
{\it (2)~Flux-qubit array.} Figure~1(b) shows an array of flux qubits. Here the small
junction in the ordinary flux qubit is replaced by a symmetric dc
SQUID to increase the tunability of the qubit. Also, a coupler
consisting of three Josephson junctions is used to produce a
controllable interqubit coupling between nearest-neighbor flux
qubits. We assume that the parameters are the same for all qubits
and also for all couplers. Moreover, the plasma frequency of
the coupler is much higher than the related qubit energy, so as to
keep the coupler in the ground state~\cite{Grajcar}. When the
leading terms are included, the Hamiltonian of the flux-qubit array
can be written as
\begin{equation}
H=\sum_{n=1}^{N-1} t\,\sigma_n^z\sigma_{n+1}^z -\sum_{n=1}^N
\left(\nu\,\sigma_n^z+\mu\,\sigma_n^x\right). \label{H-2}
\end{equation}
Here $\nu=I_p\Phi_0(\frac{1}{2}-f)$, with $I_p$ being the
persistent current of the flux qubit and $f=\Phi_q/\Phi_0+f_s/2$,
where $f_s=\Phi_s/\Phi_0$, {with $\Phi_s$ being the magnetic flux applied
in the SQUID loop [see Figure~1(d)].}
The eigenstates of the Pauli operator
$\sigma_n^z$ are the clockwise and anti-clockwise persistent-current
states of the $n$th qubit. The symmetric SQUID provides an effective
Josephson junction with coupling energy $\alpha E_J$, where
$\alpha=\cos(\pi f_s)$. The exact expression of $\mu$ in
equation~(\ref{H-2}) cannot be obtained, but it depends on $\alpha$;
numerical results~\cite{micromaser} and approximate analytical
calculations~\cite{Greenberg} showed that $\mu=0$ when $\alpha=1$.
The interqubit coupling strength reads~\cite{Grajcar}
\begin{equation}
t=\frac{\beta E_{Jc}\cos(2\pi f_c-\phi_c)}{1+2\beta\cos(2\pi
f_c-\phi_c)},
\label{T-2}
\end{equation}
where $f_c=\Phi_c/\Phi_0$ is the reduced flux applied to the coupler, and
$\phi_c=2\beta\sin(2\pi f_c)/[1+2\beta\cos(2\pi f_c)]$, {with $\beta$ being the ratio of the Josephson couplings between the smaller and larger junctions in the coupler [see Figure~1(b)].}

We study the case with $f=\frac{1}{2}$ for all flux
qubits, so as to have $\nu=0$. The Hamiltonian of the system
also becomes equation~(\ref{HF}) when applying the Jordan-Wigner
transformation:
\begin{equation}
a_n=\sigma_n^+\prod_{m=1}^{n-1}\sigma_m^x,~~~~
a_n^{\dag}=\sigma_n^-\prod_{m=1}^{n-1}\sigma_m^x,
\end{equation}
where $\sigma_n^{\pm}=\frac{1}{2}(\sigma_n^z\pm i\sigma_n^y)$.
Finally, the Hamiltonian is described by equation~(\ref{HMF}) when
introducing Majorana fermions in equation~(\ref{MF}). Therefore, the
resulting Hamiltonians in terms of Majorana fermions are the same
for both charge- and flux-qubit arrays.

For $N\rightarrow\infty$, we can obtain the energy bands of the
periodic chain by performing a Fourier transform on Hamiltonian
(\ref{HMF}):
\begin{equation}
\gamma_n^X=\sqrt{\frac{2}{N}}\sum_k e^{ikn}\gamma_k^X,
\end{equation}
where $\gamma_{-k}^X=\gamma_k^{X\dag}$ and $X=A,B$. The resulting
Hamiltonian in reciprocal space reads
\begin{equation}
H=\sum_k\left(\begin{array}{cc}\gamma_k^{A\dag}&\gamma_k^{B\dag}\end{array}\right)
\left(\begin{array}{cc}0&-iD^*(k)\\iD(k)&0\end{array}\right)
\left(\begin{array}{c}\gamma_k^A\\\gamma_k^B\end{array}\right),
\end{equation}
with $D(k)=te^{ik}+\mu$. The energy spectrum shows the particle-hole
symmetric dispersion
\begin{equation}
E(k)=\pm|D(k)|=\pm|te^{ik}+\mu|,
\label{dispersion}
\end{equation}
which consists of two bands. As examples, we present in Figure~2 the
particle-hole symmetric dispersion for $r\equiv\mu/t=0.5$ and $1$,
respectively. It is clear that when $r=1$, the gap of the two bands
closes at certain values of the wave vector $k$.

\begin{figure}
\includegraphics[width=3.3in,
bbllx=53,bblly=366,bburx=542,bbury=625]
{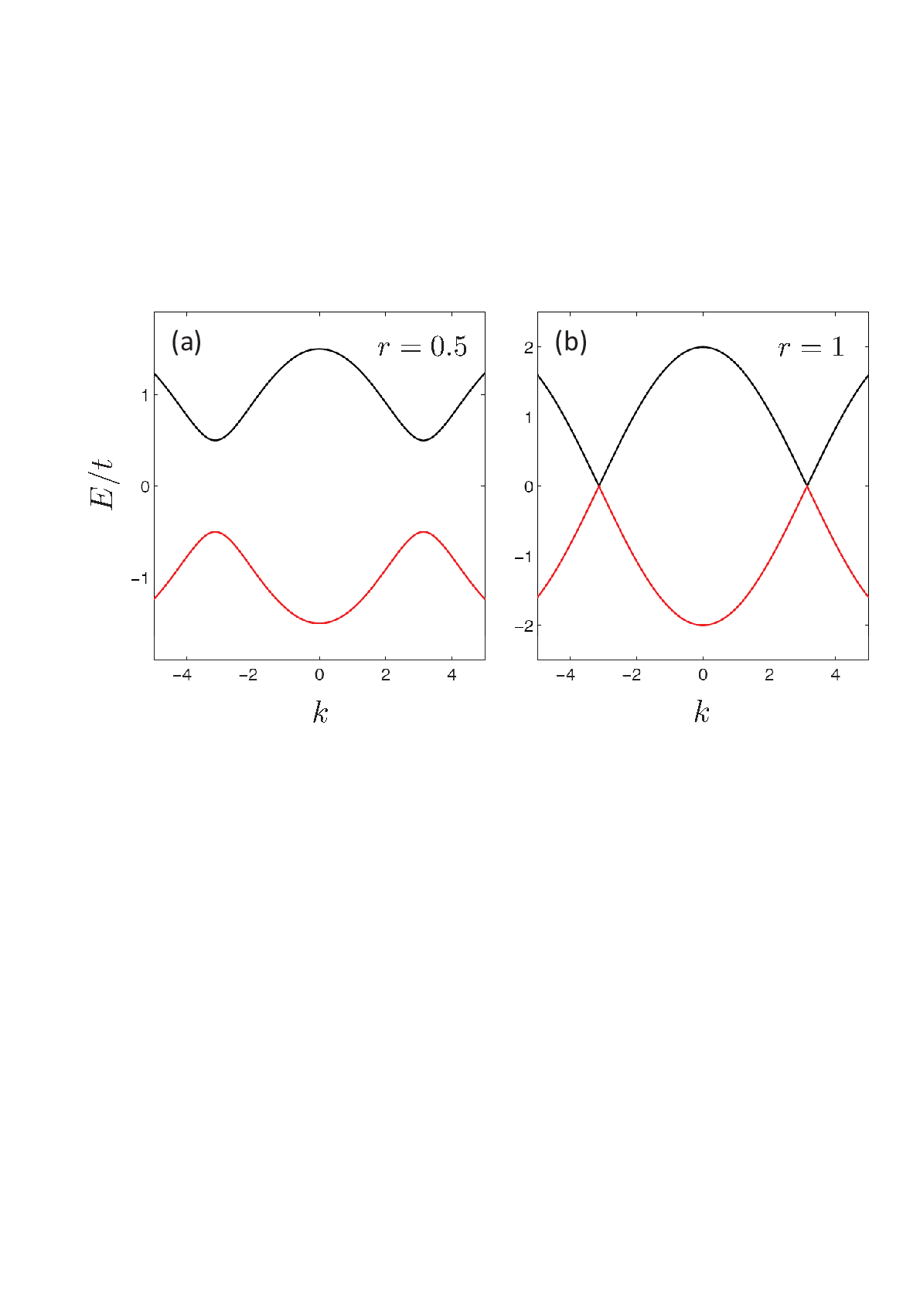} \caption{{\bf The particle-hole symmetric dispersion.}
It is obtained from equation (\ref{dispersion}), where $r\equiv\mu/t=$ (a)~$0.5$ and
(b)~$1$.} 
\label{fig2}
\end{figure}

For a finite chain, when $|r|<1$, there are two degenerate edge
modes with zero energy (i.e., in the middle of the energy gap).
These two edge modes can be represented by
\begin{equation}
Q=c_1\gamma_1^A+c_2\gamma_1^B+\cdots+c_{2N-1}\gamma_N^A
+c_{2N}\gamma_N^B,
\end{equation}
with the coefficients determined by
\begin{equation}
\mu c_{2n-1}+tc_{2n+1}=0,~~~~ tc_{2n-2}+\mu c_{2n}=0,
\end{equation}
where $n=1,2,\dots,N$, and the initial condition is $c_0=0$ for the
left-end edge state and $c_{2N+1}=0$ for the right-end edge state.
It can be derived that the left- and right-end edge modes are given,
respectively, by 
\begin{eqnarray} Q_L \!&\!=\!&\! C[\gamma_1^A-r
\gamma_2^A+r^2\gamma_3^A-\cdots+(-r)^{N-1}\gamma_N^A],\nonumber\\
Q_R \!&\!=\!&\! C[(-r)^{N-1}\gamma_1^B+\cdots+r^2\gamma_{N-2}^B-r
\gamma_{N-1}^B+\gamma_N^B],\nonumber\\
\end{eqnarray}
where the normalization factor is
$C=(\sum_{n=0}^{N-1}r^{2n})^{-1/2}$. In Ref.~\cite{Levitov}, the left-
and right-end edge modes were also studied in a flux-qubit array,
but the interqubit coupling was not tunable.

In particular, when $\mu=0$, the Hamiltonian is reduced to
\begin{equation}
H=-\sum_{n=1}^{N-1}t(2d_n^{\dag}d_n-1),
\end{equation}
where $d_n=\frac{1}{2}(\gamma_{n+1}^A+i\gamma_n^B)$ is a Dirac
fermion composed of two Majoranas at adjoining sites. The edge modes
become two unpaired Majorana fermions: $Q_L=\gamma_1^A$, and $Q_R=\gamma_N^B$,
{which emerge at the left and right ends of the chain as local modes in the Majorana-fermion representation. However, as shown in the following subsection, these two Majorana modes become non-local in the spin representation.} Moreover, these two degenerate modes do not
appear in the Hamiltonian because they have zero
energy~\cite{Kitaev-1,NJP}. Now define $|F\rangle$ to be the state
in which all eigenstates of the system with $E<0$ are occupied and
those with $E\geq 0$ are empty.
When the edge modes are occupied,
$|\Psi_L\rangle=\gamma_1^A|F\rangle$ and
$|\Psi_R\rangle=\gamma_N^B|F\rangle$ are two degenerate ground
states of the system.  These two Majorana modes can be used to
represent the basis states of a qubit {called here the Majorana qubit}:
\begin{equation}
|0\rangle\equiv d_{\rm end}|F\rangle,~~~~ |1\rangle\equiv d_{\rm
end}^{\dag}|0\rangle,
\end{equation}
where
\begin{equation}
d_{\rm end}=\frac{1}{2}(\gamma_1^A+i\gamma_N^B)
\end{equation}
is a non-local Dirac fermion, and $d_{\rm end}|0\rangle=0$. A
similar Majorana qubit was also proposed in quantum wires with
spin-orbit interactions (see, e.g., Ref.~\cite{Alicea}). Such a
qubit had initially been thought of as being fully topologically
protected, but recent studies showed that it could also suffer from
decoherence caused by either coupling to the solid-state environment
(see, e.g., Ref.~\cite{deco1}) or strong renormalization by interactions
(see, e.g., Refs.~\cite{deco2,deco3,deco4}), which was often neglected.

\vspace{.3cm}
\noindent
{\bf The Majorana qubit in the spin representation.~}Below we derive the two basis states $|0\rangle\equiv d_{\rm
end}|F\rangle$ and $|1\rangle\equiv d_{\rm end}^{\dag}|0\rangle$ in
the spin representation and discuss issues regarding the quantum
coherence of this Majorana qubit.

\vspace{.2cm}
\noindent
{\it (1)~Charge-qubit array.} When $t>0$ and $\mu=\nu=0$ in equation~(\ref{H-1}), the state
$|F\rangle$ can be written, in the spin representation, as
\begin{equation} |F\rangle=\left\{\begin{array}{ll}
\sqrt{2}\;|\!\rightarrow\leftarrow\rightarrow\leftarrow\cdots\leftarrow\rangle~~~
&\textrm{if $N=$ even},\\
\sqrt{2}\;|\!\rightarrow\leftarrow\rightarrow\leftarrow\cdots\rightarrow\rangle~~~
&\textrm{if $N=$ odd},
\end{array}\right.
\end{equation}
where $N$ is the number of charge qubits in the array and
\begin{equation}
|\!\rightarrow\rangle=\frac{1}{\sqrt{2}}\left(\begin{array}{c}
1\\1\end{array}\right),~~~
|\!\leftarrow\rangle=\frac{1}{\sqrt{2}}\left(\begin{array}{c}
1\\-1\end{array}\right)
\end{equation}
are the two eigenstates of $\sigma^x$ with eigenvalues $1$ and $-1$,
respectively.

In the spin representation, $\gamma_1^A=\sigma_1^x$ and
$\gamma_N^B=iP_c\sigma_N^x$, where the string operator
\begin{equation}
P_c\equiv\prod_{m=1}^N\sigma_m^z
\end{equation}
is the parity operator associated with the $Z_2$ symmetry of the
system. {Obviously, $\gamma_N^B$ becomes a non-local chain operator in the spin representation.}
The two degenerate ground states $|\Psi_L\rangle$ and
$|\Psi_R\rangle$ can be written as
\begin{eqnarray}
|\Psi_L\rangle \!&\!=\!&\! \gamma_1^A|F\rangle=\sigma_1^x|F\rangle \nonumber\\
&\!=\!&\!\left\{\begin{array}{ll}
\sqrt{2}\;|\!\rightarrow\leftarrow\rightarrow\leftarrow\cdots\leftarrow\rangle~~
&\textrm{if $N=$ even},\\
\sqrt{2}\;|\!\rightarrow\leftarrow\rightarrow\leftarrow\cdots\rightarrow\rangle~~
&\textrm{if $N=$ odd},
\end{array}\right.~~~
\label{leftstate}
\end{eqnarray} and
\begin{eqnarray}
|\Psi_R\rangle \!&\!=\!&\! \gamma_N^B|F\rangle=iP_c\sigma_N^x|F\rangle \nonumber\\
&\!=\!&\! \left\{\begin{array}{ll}
-i\sqrt{2}\;|\!\leftarrow\rightarrow\leftarrow\rightarrow\cdots\rightarrow\rangle~~
&\textrm{if $N=$ even},\\
i\sqrt{2}\;|\!\leftarrow\rightarrow\leftarrow\rightarrow\cdots\leftarrow\rangle~~
&\textrm{if $N=$ odd}.
\end{array}\right.~~~
\label{rightstate}
\end{eqnarray}
It is clear that $\langle\Psi_L|\Psi_R\rangle=0$.
The two basis states $|0\rangle\equiv d_{\rm end}|F\rangle$
and $|1\rangle\equiv d_{\rm end}^{\dag}|0\rangle$ of the Majorana
qubit are given, respectively, by
\begin{eqnarray}
|0\rangle \!&\!=\!&\!\frac{1}{2}(\gamma_1^A+i\gamma_N^B)|F\rangle
\nonumber\\
&\!=\!&\!\left\{\begin{array}{l}
\!\!\frac{|\rightarrow\leftarrow\rightarrow\leftarrow\cdots\leftarrow\rangle
+|\leftarrow\rightarrow\leftarrow\rightarrow\cdots\rightarrow\rangle}{\sqrt{2}}~~
\textrm{if $N=$ even},\\
\!\!\frac{|\rightarrow\leftarrow\rightarrow\leftarrow\cdots\rightarrow\rangle
-|\leftarrow\rightarrow\leftarrow\rightarrow\cdots\leftarrow\rangle}{\sqrt{2}}~~
\textrm{if $N=$ odd},
\end{array}\right.~~~
\label{state-0}
\end{eqnarray}
and
\begin{eqnarray}
|1\rangle
\!&\!=\!&\!\frac{1}{2}(\gamma_1^A-i\gamma_N^B)|0\rangle=\frac{1}{2}(\sigma_1^x+P_c\sigma_N^x)|0\rangle
\nonumber\\
&\!=\!&\!\left\{\begin{array}{l}
\!\!\frac{|\rightarrow\leftarrow\rightarrow\leftarrow\cdots\leftarrow\rangle
-|\leftarrow\rightarrow\leftarrow\rightarrow\cdots\rightarrow\rangle}{\sqrt{2}}~~
\textrm{if $N=$ even},\\
\!\!\frac{|\rightarrow\leftarrow\rightarrow\leftarrow\cdots\rightarrow\rangle
+|\leftarrow\rightarrow\leftarrow\rightarrow\cdots\leftarrow\rangle}{\sqrt{2}}~~
\textrm{if $N=$ odd}.
\end{array}\right.~~~
\label{state-1}
\end{eqnarray}
{Note that $|1\rangle\equiv d_{\rm end}^{\dag}|0\rangle$ can also be written as
$|1\rangle=d_{\rm end}^{\dag}d_{\rm end}|F\rangle=\frac{1}{2}(1+i\gamma_1^A\gamma_N^B)|F\rangle$,
which is identical to equation (\ref{state-1}).}
These two basis states of the Majorana qubit are also two degenerate
ground states of the system. Moreover, these ground states have
well-defined parities because $P_c|0\rangle=|0\rangle$ and
$P_c|1\rangle=-|1\rangle$ if $N=\textrm{even}$, and because
$P_c|0\rangle=-|0\rangle$ and $P_c|1\rangle=|1\rangle$ if
$N=\textrm{odd}$. In Ref.~\cite{Tserk}, similar Majorana modes in spin-chain networks were also
used to encode a qubit.

When the externally-tunable parameters such as gate voltages and
applied fluxes are identified at each qubit, equation~(\ref{H-1})
can be rewritten as
\begin{equation}
H=\sum_{n=1}^{N-1} t_{n,n+1}\,\sigma_n^x\sigma_{n+1}^x -\sum_{n=1}^N
\left(\mu_n\,\sigma_n^z+\nu_n\,\sigma_n^x\right),
\end{equation}
where
\begin{equation}
\mu_n=\frac{1}{2}E_{\rm
ch}\left(1-\frac{C_gV_g^{(n)}}{e}\right),~~
\nu_n=E_{J0}\cos\left(\frac{\pi\Phi_q^{(n)}}{\Phi_0}\right),
\label{parameter-1}
\end{equation}
and the interqubit coupling is given by
\begin{equation}
t_{n,n+1}=L_J\left(\frac{\pi E_{J0}}{\Phi_0}\right)^2
\sin\left(\frac{\pi\Phi_q^{(n)}}{\Phi_0}\right)
\sin\left(\frac{\pi\Phi_q^{(n+1)}}{\Phi_0}\right).
\label{parameter-2}
\end{equation}
As noted in Ref.~\cite{Kitaev-2}, if $\sum_{n=1}^N\nu_n\sigma_n^x\neq
0$, this longitudinal term will lift the state degeneracy of the
system. However, in our designed circuits, this can be avoided
because we can have $\nu_n=0$ for each qubit by tuning the external
flux to $\Phi_q^{(n)}=\frac{1}{2}\Phi_0$. Also, we can have
$\mu_n=0$ by tuning the gate voltage to $V_g^{(n)}=e/C_g$, so as to
achieve unpaired Majorana modes {emerging at the two ends of the charge-qubit
array in the Majorana-fermion representation.}

{With regard to the quantum coherence of the Majorana qubit, there are
three types of local perturbations that we should
consider:~(i)~$\delta\mu_n\,\sigma_n^z$, (ii)~$\delta
\nu_n\,\sigma_n^x$, and (iii)~$\delta t_{n,n+1}\,\sigma_n^x\sigma_{n+1}^x$.
The charge perturbation $\delta\mu_n\,\sigma_n^z$ can be explicitly written as}
\begin{equation}
\delta\mu_n=-\left(\frac{E_{\rm ch}}{2e}\right)\delta Q_n,
\end{equation}
where
\begin{equation}
\delta Q_n=C_g\delta V_g^{(n)}+\delta Q_b^{(n)},
\end{equation}
{with the term $C_g\delta V_g^{(n)}$ arising from the gate-voltage fluctuations and $\delta Q_b^{(n)}$ being due to the background charge fluctuations (e.g., the two-level fluctuators). As shown in equations (\ref{parameter-1}) and (\ref{parameter-2}), the parameters $\nu_n$ and $t_{n,n+1}$ contain both the Josephson coupling $E_{J0}$ and the flux $\Phi_q^{(n)}$. Therefore, the local perturbations $\delta\nu_n\,\sigma_n^x$ and $\delta t_{n,n+1}\,\sigma_n^x\sigma_{n+1}^x$ can be contributed by both the critical-current~\cite{Palmer} and flux fluctuations.}

The local perturbation $\delta\mu_n\,\sigma_n^z$ can only tend to
drive the ground state (i.e., the Majorana-qubit state) $|0\rangle$
($|1\rangle$) to an excited state, which has an energy level higher
than the ground state. This is owing to the protection of the
Majorana-mode states $|\Psi_L\rangle$ and $|\Psi_R\rangle$ against
the local perturbation $\delta\mu_n\,\sigma_n^z$, because this
perturbation cannot produce a state transition (relaxation) between
$|\Psi_L\rangle$ and $|\Psi_R\rangle$. Actually, the local
perturbation $\delta\mu_n\,\sigma_n^z$ tends to drive
$|\Psi_L\rangle$ ($|\Psi_R\rangle$) to an excited state with an
energy difference $\Delta$ from the ground state, where $\Delta=4t$
for $1<n<N$, and $\Delta=2t$ for $n=1$ and $N$. Nevertheless, such a
state transition is not permitted for a small perturbation
$\delta\mu_n\,\sigma_n^z$. Thus, {the local perturbation
$\delta\mu_n\,\sigma_n^z$ (i.e., the charge fluctuations) will not produce decoherence to the
Majorana qubit. This is a distinct advantage of the Majorana qubit over a single charge qubit in which the charge fluctuations dominate.}

{The environmentally-induced decoherence in a near-critical 1D system of $N\gg 1$ coupled qubits was studied in Ref.~\cite{Khvesh}, where a model Hamiltonian analogous to equation (\ref{H-1}) with $\nu=0$ was used and only the local magnetic-field fluctuations (i.e., the local perturbation $\delta\mu_n\,\sigma_n^z$ in our model) were considered. It was found that the requirement of preserving the qubits' entanglement over a certain idling time between consecutive gates can be better fulfilled away from criticality, i.e., when $r(\equiv\mu/t)\neq 1$. In our study, we consider the case with $r=0$, which is away from the criticality, and the two Majorana-qubit states $|0\rangle$ and $|1\rangle$ are entangled states of multiple qubits [see equations (\ref{state-0}) and (\ref{state-1})]. Indeed, as discussed above, these multi-qubit entangled states are robust against the local perturbation $\delta\mu_n\,\sigma_n^z$.
}

{As for the local perturbations $\delta\nu_n\,\sigma_n^x$
and $\delta t_{n,n+1}\,\sigma_n^x\sigma_{n+1}^x$, they should randomly shift the energy levels of the states
$|\Psi_L\rangle$ and $|\Psi_R\rangle$, causing pure dephasing to these Majorana-mode states. However, while $\delta t_{n,n+1}\,\sigma_n^x\sigma_{n+1}^x$ yields pure dephasing to the Majorana-qubit states $|0\rangle$ and $|1\rangle$, the
local perturbation $\delta\nu_n\,\sigma_n^x$ produces relaxation to these Majorana-qubit states.
In a circuit composed of inductively-coupled charge qubits, the interqubit coupling is usually much smaller than the Josephson coupling energy $E_{J0}$,
so the coupler perturbation $\delta t_{n,n+1}\,\sigma_n^x\sigma_{n+1}^x$ should be weaker than $\delta\nu_n\,\sigma_n^x$. As shown above, the Majorana qubit is robust again the charge noise. Now, the dominant noise in the Majorana qubit is due to the perturbation $\delta\nu_n\,\sigma_n^x$ involving both critical-current and flux fluctuations.
In order to have a longer decoherence time, a single charge qubit
usually works at the optimal (i.e., degeneracy) point $N_g\equiv eV_g/2e=1/2$.
When this single charge qubit is slightly away from the optimal charge degeneracy point,
the decoherence time becomes drastically short because of its strong sensitivity to the charge
noise. Nevertheless, the Majorana qubit consisting
of a charge-qubit array is robust against the charge noise. Then, its quantum
coherence is still preserved even if each charge qubit is randomly
shifted away from the optimal charge degeneracy point. This is also
one of the advantages of the Majorana qubit over a single charge
qubit.}

{In Ref.~\cite{Pedrocchi}, an inhomogeneous spin ladder was proposed to study the robustness of the Majorana modes. This spin model is an inhomogeneous ladder version of the Kitaev honeycomb model~\cite{Kitaev-3}. Similar to the 1D quantum Ising model, the zero-energy Majorana modes of the inhomogeneous spin ladder are also localized in the fermionic representation and emerge at either the two ends of the ladder or the boundary between sections in different topological phases~\cite{Pedrocchi}. As shown above, in
the quantum Ising model described by equation (\ref{H-1}), the topological ground-state degeneracy is robust against the local perturbation $\delta\mu_n\,\sigma_n^z$, but can be lifted by the local perturbation $\delta\nu_n\,\sigma_n^x$. In the inhomogeneous spin ladder, the topological ground-state degeneracy cannot be fully lifted by inhomogeneous magnetic fields purely along the $x$, $y$ or $z$ direction~\cite{Pedrocchi}. This is the advantage of the inhomogeneous spin ladder. However, as further shown in Ref.~\cite{Pedrocchi}, the topological ground-state degeneracy of the inhomogeneous spin ladder can be lifted by local two-body terms. In the 1D quantum Ising model in equation (\ref{H-1}), the two-body (i.e., coupler) perturbation $\delta t_{n,n+1}\,\sigma_n^x\sigma_{n+1}^x$ can also lift the topological ground-state degeneracy, but compared with the local perturbation $\delta\mu_n\,\sigma_n^z$, the local perturbation $\delta\nu_n\,\sigma_n^x$ is much weaker and the coupler perturbation $\delta t_{n,n+1}\,\sigma_n^x\sigma_{n+1}^x$ is even weaker in the 1D quantum Ising model realized using a charge-qubit array.
}

{As a variation of the charge qubit, the transmon qubit was also often used in superconducting
quantum circuits~\cite{transmon}. In this qubit, the perturbation $\delta
\nu_n\,\sigma_n^x$, which can be due to the fluctuations of flux, cavity photons and critical current,
is more important than the perturbation $\delta\mu_n\,\sigma_n^z$ arising from the charge noise.
In the Majorana qubit with charge qubits replaced by transmons, the perturbation $\delta\nu_n\,\sigma_n^x$ becomes more important, but the advantage of the Majorana qubit regarding the insensitivity to the charge noise will still remain. Therefore, the quantum coherence of the Majorana qubit is preserved even if each transmon shifts randomly
away from the optimal charge degeneracy point.}

\vspace{.2cm}
\noindent
{\it (2)~Flux-qubit array.} When $t>0$ and $\mu=\nu=0$ in equation~(\ref{H-2}), the
state $|F\rangle$ can be written, in the spin representation, as
\begin{equation}
|F\rangle=\left\{\begin{array}{ll}
\sqrt{2}\;|\!\uparrow\downarrow\uparrow\downarrow\cdots\downarrow\rangle~~~
&\textrm{if $N=$ even},\\
\sqrt{2}\;|\!\uparrow\downarrow\uparrow\downarrow\cdots\uparrow\rangle~~~
&\textrm{if $N=$ odd},
\end{array}\right.
\end{equation}
where
\begin{equation}
|\!\uparrow\rangle=\left(\begin{array}{c} 1\\0\end{array}\right),~~~
|\!\downarrow\rangle=\left(\begin{array}{c} 0\\1\end{array}\right)
\end{equation}
are the two eigenstates of $\sigma^z$ with eigenvalues $1$ and $-1$,
respectively.

It can be derived that $\gamma_1^A=\sigma_1^z$ and
$\gamma_N^B=-i\sigma_N^zP_f$, where the parity operator associated
with the $Z_2$ symmetry of the system is given by
\begin{equation}
P_f\equiv\prod_{m=1}^N\sigma_m^x.
\end{equation}
In the spin representation, the two degenerate ground states
$|\Psi_L\rangle$ and $|\Psi_R\rangle$ can be written as
\begin{eqnarray}
|\Psi_L\rangle \!&\!=\!&\!\gamma_1^A|F\rangle=\sigma_1^z|F\rangle \nonumber\\
&\!=\!&\!\left\{\begin{array}{ll}
\sqrt{2}\;|\!\uparrow\downarrow\uparrow\downarrow\cdots\downarrow\rangle~~
&\textrm{if $N=$ even},\\
\sqrt{2}\;|\!\uparrow\downarrow\uparrow\downarrow\cdots\uparrow\rangle~~
&\textrm{if $N=$ odd},~~~~
\end{array}\right.
\end{eqnarray}
and
\begin{eqnarray}
|\Psi_R\rangle \!&\!=\!&\!\gamma_N^B|F\rangle=-i\sigma_N^zP_f|F\rangle \nonumber\\
&\!=\!&\!\left\{\begin{array}{ll}
-i\sqrt{2}\;|\!\downarrow\uparrow\downarrow\uparrow\cdots\uparrow\rangle~~
&\textrm{if $N=$ even},\\
i\sqrt{2}\;|\!\downarrow\uparrow\downarrow\uparrow\cdots\downarrow\rangle~~
&\textrm{if $N=$ odd}.
\end{array}\right.
\end{eqnarray}
{Also, it is clear that $\langle\Psi_L|\Psi_R\rangle=0$.}
The two basis states $|0\rangle\equiv d_{\rm end}|F\rangle$ and
$|1\rangle\equiv d_{\rm end}^{\dag}|0\rangle$ of the Majorana qubit
are given, respectively, by
\begin{eqnarray}
|0\rangle \!&\!=\!&\!\frac{1}{2}(\gamma_1^A+i\gamma_N^B)|F\rangle
\nonumber\\
&\!=\!&\!\left\{\begin{array}{l}
\!\!\frac{|\uparrow\downarrow\uparrow\downarrow\cdots\downarrow\rangle
+|\downarrow\uparrow\downarrow\uparrow\cdots\uparrow\rangle}{\sqrt{2}}~~~
\textrm{if $N=$ even},\\
\!\!\frac{|\uparrow\downarrow\uparrow\downarrow\cdots\uparrow\rangle
-|\downarrow\uparrow\downarrow\uparrow\cdots\downarrow\rangle}{\sqrt{2}}~~~
\textrm{if $N=$ odd},
\end{array}\right.
\end{eqnarray}
and
\begin{eqnarray}
|1\rangle
\!&\!=\!&\!\frac{1}{2}(\gamma_1^A-i\gamma_N^B)|0\rangle=\frac{1}{2}(\sigma_1^z-\sigma_N^zP_f)|0\rangle
\nonumber\\
&\!=\!&\!\left\{\begin{array}{l}
\!\!\frac{1|\uparrow\downarrow\uparrow\downarrow\cdots\downarrow\rangle
-|\downarrow\uparrow\downarrow\uparrow\cdots\uparrow\rangle}{\sqrt{2}}~~~
\textrm{if $N=$ even},\\
\!\!\frac{|\uparrow\downarrow\uparrow\downarrow\cdots\uparrow\rangle
+|\downarrow\uparrow\downarrow\uparrow\cdots\downarrow\rangle}{\sqrt{2}}~~~
\textrm{if $N=$ odd},
\end{array}\right.
\end{eqnarray}
These two basis states of the Majorana qubit are also two degenerate
ground states of the system and have well-defined parities.

When the parameters are identified at each qubit, we can rewrite equation~(\ref{H-2}) as
\begin{equation}
H=\sum_{n=1}^{N-1} t_{n,n+1}\,\sigma_n^z\sigma_{n+1}^z -\sum_{n=1}^N
\left(\nu_n\,\sigma_n^z+\mu_n\,\sigma_n^x\right).
\end{equation}
Here $\nu_n=I_p\Phi_0(\frac{1}{2}-f_n)$, and
$f_n=\Phi_q^{(n)}/\Phi_0+f_s^{(n)}/2$, where
$f_s^{(n)}=\Phi_s^{(n)}/\Phi_0$. The interqubit coupling reads
\begin{equation}
t_{n,n+1}=\frac{\beta E_{Jc}\cos(2\pi
f_c^{(n,n+1)}-\phi_c^{(n,n+1)})}{1+2\beta\cos(2\pi
f_c^{(n,n+1)}-\phi_c^{(n,n+1)})},
\end{equation}
where
\begin{equation}
\phi_c^{(n,n+1)}=\frac{2\beta\sin(2\pi
f_c^{(n,n+1)})}{1+2\beta\cos(2\pi f_c^{(n,n+1)})},
\end{equation}
and $f_c^{(n,n+1)}=\Phi_c^{(n,n+1)}/\Phi_0$ is the reduced flux
applied to the coupler between qubits $n$ and $n+1$.

The local perturbation $\delta\mu_n\,\sigma_n^x$ can only tend to
drive the ground state (i.e., the Majorana-qubit state) $|0\rangle$
($|1\rangle$) to an excited state of the system which has an energy
level higher than the ground state. Similar to the case of the
charge-qubit array, this is also owing to the protection of the
Majorana-mode states $|\Psi_L\rangle$ and $|\Psi_R\rangle$ against
the local perturbation $\delta\mu_n\,\sigma_n^x$. Indeed, the local
perturbation $\delta\mu_n\,\sigma_n^x$ tends to drive
$|\Psi_L\rangle$ ($|\Psi_R\rangle$) to an excited state with an
energy difference $\Delta$ from the ground state, where $\Delta=4t$
for $1<n<N$, and $\Delta=2t$ for $n=1$ and $N$. Nevertheless, such a
state transition is not permitted for a small perturbation
$\delta\mu_n\,\sigma_n^x$. Therefore, in contrast to a single flux
qubit, the local perturbation $\delta\mu_n\,\sigma_n^x$ will not
produce decoherence to the Majorana qubit.

{The local perturbation $\delta\nu_n\,\sigma_n^z$ randomly shifts the energy levels
of the Majorana-mode states $|\Psi_L\rangle$ and $|\Psi_R\rangle$
to cause pure dephasing to these states. Also, it
produces relaxation to the Majorana-qubit states $|0\rangle$ and
$|1\rangle$. However, the coupler perturbation $\delta t_{n,n+1}\,\sigma_n^z\sigma_{n+1}^z$
yields pure dephasing to both the Majorana-mode states ($|\Psi_L\rangle$
and $|\Psi_R\rangle$) and the Majorana-qubit states ($|0\rangle$ and
$|1\rangle$). Because the interqubit coupling is usually much smaller than $I_p\Phi_0$,
the coupler perturbation $\delta t_{n,n+1}\,\sigma_n^z\sigma_{n+1}^z$ should be much weaker than
$\delta\nu_n\,\sigma_n^z$. Therefore, in the case of a flux-qubit array, the dominant noise of the Majorana qubit
is due to the perturbation $\delta\nu_n\,\sigma_n^z$. In order to improve the quantum
coherence of the Majorana qubit, one can suppress the fluctuations
$\delta\nu_n$ by reducing $I_p$.}
This can be achieved by
reducing the size of the Josephson junctions in each flux qubit
because $I_p$ is proportional to the Josephson coupling energy
$E_J$. Note that when reducing the size of the Josephson junctions to
suppress flux noise, the charge noise can finally become important,
due to the increasing charging energy. Thus, while the flux noise is
suppressed, one can shunt a large capacitance to the Josephson
junction, so as to suppress the charge noise as well. This method
was proposed to increase the decoherence time of the flux
qubit~\cite{low-decoherence} and has been implemented in a recent
experiment~\cite{enhanced-qubit}.

\vspace{.2cm}
\noindent
{\bf Manipulating and probing Majorana modes.~}The superconducting-qubit arrays proposed above can be used
to realize a tunable 1D quantum Ising model on wire networks,
similar to the semiconducting wire networks in Ref.~\cite{Alicea}, to
demonstrate the non-Abelian statistics of Majorana fermions. In
particular, braiding Majoranas can be implemented via a T-junction
formed by two perpendicular wires~\cite{Alicea}. Here we use four
superconducting qubits, as the smallest size of the system, to form
such a T-junction [see Figure~3(a)], where $\nu=0$ for all charge (flux) qubits.
When the Jordan-Wigner tranformation is performed, this T-junction of four
qubits is described by
\begin{eqnarray}
H\!&\!=\!&\!t(a_1-a_1^{\dag})(a_2^{\dag}+a_2)+t(a_{1'}
-a_{1'}^{\dag})(a_2^{\dag}+a_2)\nonumber\\
&&\!+t(a_2-a_2^{\dag})(a_3^{\dag}+a_3)-\mu(2a_1^{\dag}a_1-1)\nonumber\\
&&\!-\mu(2a_{1'}^{\dag}a_{1'}-1)-\mu(2a_2^{\dag}a_2-1)
-\mu(2a_3^{\dag}a_3-1)\nonumber\\
&\!=\!&\!i(t\gamma_1^B\gamma_2^A+t\gamma_{1'}^B\gamma_2^A
+t\gamma_2^B\gamma_3^A)\nonumber\\
&&\!-i(\mu\gamma_1^A\gamma_1^B+\mu\gamma_{1'}^A\gamma_{1'}^B
+\mu\gamma_2^A\gamma_2^B+\mu\gamma_3^A\gamma_3^B), \label{H-T}
\end{eqnarray}
where qubits are numbered by starting from sites 1 and $1'$ and
ending at site 3.

\begin{figure}
\includegraphics[width=3.3in,
bbllx=32,bblly=128,bburx=561,bbury=723]
{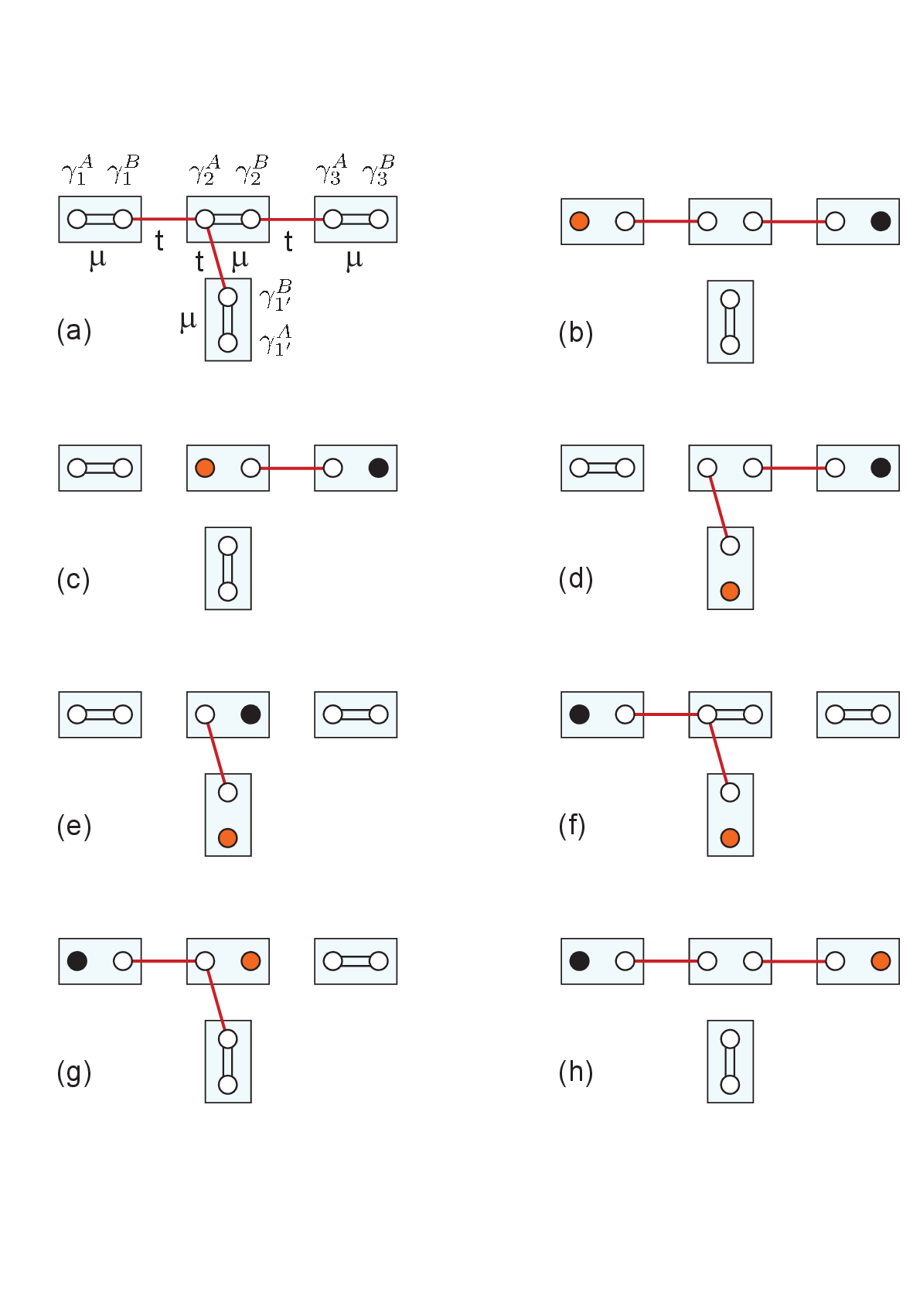}\caption{{\bf Braiding two unpaired Majorana fermions.} (a)~T-junction
formed by qubits 1, 2, 3 and $1'$, where each qubit is denoted by a
rectangular box. Two Majoranas related to the same qubit (e.g.,
$\gamma_1^A$ and $\gamma_1^B$) can be paired by the parameter $\mu$
while two Majoranas related to adjoining qubits (e.g., $\gamma_1^B$
and $\gamma_2^A$) can be paired by $t$. (b)~Unpaired-Majorana region
for the whole horizontal array, where an unpaired Majorana (denoted
by a solid circle) is located at each end. (c)~Adiabatically tuning
$\mu$ to nonzero for qubit 1 and turning off $t$ between qubits 1
and 2 drive the left-end Majorana mode (shown in orange) to the
middle qubit. (d)~Adiabatically tuning $\mu$ to zero for qubit $1'$
and turning on $t$ between qubits 2 and $1'$ drive the original
left-end Majorana to the bottom of the T-junction. (e)~The right-end
Majorana (in black) is driven to the middle qubit by adiabatically
tuning $\mu$ to nonzero for qubit 3 and turning off $t$ between
qubits 2 and 3. (f)~The original right-end Majorana is finally
driven to the left end by adiabatically tuning $\mu$ to a
sufficiently large value for qubit 2, tuning $\mu$ to zero for qubit
1 and turning on $t$ between qubits 1 and 2. (g)~The Majorana at the
bottom is driven to the middle qubit by adiabatically tuning $\mu$
to zero for qubit 2 and tuning $\mu$ to a sufficiently large value
for qubit $1'$. (h)~Adiabatically turning off $t$ between qubits 2
and $1'$, tuning $\mu$ to zero for qubit 3, and turning on $t$
between qubits 2 and 3 finally drive the original left-end Majorana
to the right end. This accomplishes the anti-clockwise braiding of
two Majorana fermions.}
\label{fig3}
\end{figure}

For the Hamiltonian (\ref{H-T}), when $t=0$ for all pairs of
{adjoining qubits, the nearest-neighbor Majoranas related to the same site are coupled by $\mu$, and the nearest-neighbor Majoranas related to two adjoining sites are decoupled. Then, Hamiltonian (\ref{H-T}) is reduced to}
\begin{equation}
H=-i(\mu\gamma_1^A\gamma_1^B+\mu\gamma_{1'}^A\gamma_{1'}^B
+\mu\gamma_2^A\gamma_2^B+\mu\gamma_3^A\gamma_3^B).
\label{H-T-1}
\end{equation}
In this case, the Majoranas are all paired in the whole
T-junction region and no edge states occur. Starting from this
phase, we adiabatically vary the parameters of superconducting
qubits to have the horizontal array become an unpaired-Majarana
region, i.e., {adiabatically turn the parameter $\mu$ to zero for each qubit in the horizontal array and simultaneously switch on the interqubit coupling $t$ for the horizontal array. Then, the Hamiltonian (\ref{H-T-1}) becomes}
\begin{equation}
H=i(t\gamma_1^B\gamma_2^A+t\gamma_2^B\gamma_3^A)
-i\mu\gamma_{1'}^A\gamma_{1'}^B.
\label{H-T-2}
\end{equation}
{This corresponds to the configuration of Majoranas in Figure~3(b), where a pair of isolated Majoranas emerge at the two ends of the horizontal array.} Here adiabatic changes of the parameters
with respect to the time $\tau$ require that
\begin{equation}
\hbar|d\mu/d\tau|\ll E_g^2,~~~~\hbar|dt/d\tau|\ll E_g^2,
\end{equation}
where $E_g$ is the energy gap between the first excited and ground
states of the system at the time $\tau$. Generally, the system takes
the superposition state of these two degenerate Majorana modes:
$|\Psi\rangle=(a\gamma_1^A+b\gamma_3^B)|F\rangle$, where $|F\rangle$
is the state in which all eigenstates of the system with $E<0$ are
occupied. However, while reaching the state in Figure~3(b),
if $\mu$ for qubits 1, 2 and 3 are all adiabatically tuned to zero
in the same manner and the interqubit coupling between qubits 1 and
2 is adiabatically switched on in the same way as that between
qubits 2 and 3, then the left- and right-end Majoranas should occur
with equal probabilities. Using this state
$|\Psi\rangle_i=\frac{1}{2}(\gamma_1^A+e^{i\theta}\gamma_3^B)|F\rangle$
as the initial state, one can braid the left- and right-end
Majoranas through the steps shown in Figures~3(c)-3(h) by adiabatically tuning the
qubit parameters. {For instance, by adiabatically switching on $\mu$ for qubit 1 and turning off the coupling $t$ between qubits 1 and 2, Hamiltonian (\ref{H-T-2}) is changed to}
\begin{equation}
H=-i(\mu\gamma_{1}^A\gamma_{1}^B+\mu\gamma_{1'}^A\gamma_{1'}^B)+it\gamma_2^B\gamma_3^A,
\end{equation}
{i.e., the configuration of Majoranas in Figure~3(b) is adiabatically converted to the configuration of Majoranas in Figure~3(c). Similarly, other steps shown in Figures~3(d)-3(h) can be achieved.}
This braiding of Majoranas following the steps from Figure~3(b) to 3(h) corresponds to a
unitary operator~\cite{Ivanov,Alicea}
which transforms $\gamma_1^A$ to 
$\gamma_3^B$ and $\gamma_3^B$ to 
$-\gamma_1^A$.
Therefore, the initial state $|\Psi\rangle_i$ of the system is
transferred to
$|\Psi\rangle_f=\frac{1}{2}(\gamma_3^B-e^{i\theta}\gamma_1^A)|F\rangle$
after braiding the left- and right-end Majoranas.

Finally, we focus on probing Majorana fermions. The initial state
$|\Psi\rangle_i=\frac{1}{2}(\gamma_1^A+e^{i\theta}\gamma_3^B)|F\rangle$
given in Figure~3(b) is a ground state of the system with
qubit $1'$ decoupled from the horizontal array of superconducting
qubits. When expressed in the basis states of qubits, this initial
state can be written as
$|\Psi\rangle_i=|\Psi_{123}\rangle_i\otimes|\Psi_{1'}\rangle_i$,
where
\begin{eqnarray}
|\Psi_{123}\rangle_i\!&\!=\!&\!\lambda_{i1}|0_10_20_3\rangle
+\lambda_{i2}|0_10_21_3\rangle+\lambda_{i3}|0_11_20_3\rangle\nonumber\\
&&\!+\lambda_{i4}|0_11_21_3\rangle+\lambda_{i5}|1_10_20_3\rangle
+\lambda_{i6}|1_10_21_3\rangle\nonumber\\
&&\!+\lambda_{i7}|1_11_20_3\rangle+\lambda_{i8}|1_11_21_3\rangle,
\end{eqnarray}
and
$|\Psi_{1'}\rangle_i=\xi_{i1}|0_{1'}\rangle+\xi_{i2}|1_{1'}\rangle$.
Also, the final state
$|\Psi\rangle_f=\frac{1}{2}(\gamma_3^B-e^{i\theta}\gamma_1^A)|F\rangle$
is another degenerate ground state of the same system and can be
expressed as $|\Psi\rangle_f=|\Psi_{123}\rangle_f\otimes
|\Psi_{1'}\rangle_f$, where
$|\Psi_{1'}\rangle_f=\xi_{f1}|0_{1'}\rangle+\xi_{f2}|1_{1'}\rangle$,
and $|\Psi_{123}\rangle_f$ has the same form as
$|\Psi_{123}\rangle_i$, but the $\lambda_{il}$ are replaced by
$\lambda_{fl}$, with $l=1$ to 8. The states $|\Psi\rangle_i$ and
$|\Psi\rangle_f$ can be distinguished using experimentally available
state-tomography techniques for superconducting qubits (see, e.g.,
Refs.~~\cite{Steffen,Filipp}), which involve reconstructing an unknown
quantum state from a complete set of measurements of the system
observables. For the initial state in Figure~3(b) and the final
state in Figure~3(h), the qubit $1'$ is decoupled from the array
consisting of the three coupled qubits 1, 2 and 3. Thus, only
one-qubit tomography for qubit $1'$ and three-qubit tomography for
coupled qubits 1, 2 and 3 are required for distinguishing the
initial and final states.

{Note that the initial and final states after braiding $\gamma_1^A$ and $\gamma_3^B$ can be written as
$|\Psi\rangle_i=\frac{1}{2}(\gamma_1^A|F\rangle+e^{i\theta}\gamma_3^B|F\rangle)$ and
$|\Psi\rangle_f=\frac{1}{2}(\gamma_3^B|F\rangle-e^{i\theta}\gamma_1^A|F\rangle)$, which have different relative phases between $\gamma_1^A|F\rangle$ and $\gamma_3^B|F\rangle$. Because the qubit $1'$ is decoupled from the horizontal array consisting of the three coupled qubits 1, 2 and 3, the states $|\Psi_{1'}\rangle_i$ and $|\Psi_{1'}\rangle_f$ are the same, in addition to a global phase between them. Thus, when performing quantum-state tomography, the different relative phases between $\gamma_1^A|F\rangle$ and $\gamma_3^B|F\rangle$ will give rise to the difference between $|\Psi_{123}\rangle_i$ and $|\Psi_{123}\rangle_f$.}

Experimentally, it is more complicated to use state-tomography
techniques to determine the quantum state of three qubits other than
two qubits. Therefore, we can consider the state
$|\bar{\Psi}\rangle_i$ in Figure~3(c) as the initial state.
This state is a ground state of the system with qubits $1'$ and 1
decoupled from other qubits and can be decomposed as
$|\bar{\Psi}\rangle_i=|\Psi_{23}\rangle_i\otimes|\Psi_{1'}\rangle_i\otimes|\Psi_1\rangle_i$,
where
\begin{eqnarray}
|\Psi_{23}\rangle\!&\!=\!&\!\lambda_{i1}|0_20_3\rangle
+\lambda_{i2}|0_21_3\rangle+\lambda_{i3}|1_20_3\rangle+\lambda_{i4}|1_21_3\rangle,\nonumber\\
|\Psi_{1'}\rangle_i\!&\!=\!&\!\xi_{i1}|0_{1'}\rangle+\xi_{i2}|1_{1'}\rangle,~~~
|\Psi_1\rangle_i=\eta_{i1}|0_1\rangle+\eta_{i2}|1_1\rangle.\nonumber\\
\end{eqnarray}
From the state in Figure~3(h), further proceeding with one
step analogous to that from Figure~3(b) to
Figure~3(c), we achieve the final state with the originally
unpaired Majoranas $\gamma_2^A$ and $\gamma_3^B$ braided. This final
state can also be decomposed as
$|\bar{\Psi}\rangle_f=|\Psi_{23}\rangle_f\otimes
|\Psi_{1'}\rangle_f\otimes|\Psi_1\rangle_f$, where
$|\Psi_{1'}\rangle_f=\xi_{f1}|0_{1'}\rangle+\xi_{f2}|1_{1'}\rangle$,
$|\Psi_1\rangle_f=\eta_{f1}|0_1\rangle+\eta_{f2}|1_1\rangle$, and
$|\Psi_{23}\rangle_f$ has the same form as $|\Psi_{23}\rangle_i$,
but the $\lambda_{il}$ are replaced by $\lambda_{fl}$, with $l=1$ to
4. Similarly, the states $|\bar{\Psi}\rangle_i$ and
$|\bar{\Psi}\rangle_f$ can also be discriminated using
state-tomography techniques. Here, because qubits 1 and $1'$ are
decoupled from the two coupled qubits 2 and 3, only two-qubit
tomography for coupled qubits 2 and 3 as well as one-qubit
tomography for qubits 1 and $1'$ are needed for distinguishing the
initial and final states.

Experimentally, in addition to one-qubit
tomography, two-qubit tomography is also implementable for
superconducting qubits (see, e.g., Refs.~\cite{Steffen,Filipp}). {Thus, it is feasible to measure $|\bar{\Psi}\rangle_i$ and $|\bar{\Psi}\rangle_f$ because the two-qubit tomography can be used to determine $|\Psi_{23}\rangle_i$ and $|\Psi_{23}\rangle_f$. Moreover, quantum-state tomography has been performed on three~\cite{DiCarlo} or even five~\cite{Barends} superconducting qubits, so it also becomes feasible to measure $|\Psi\rangle_i$ and $|\Psi\rangle_f$ by determining $|\Psi_{123}\rangle_i$ and $|\Psi_{123}\rangle_f$ via quantum-state tomography. This is important here since information might be lost by only performing two-qubit tomography, particularly in the case of poor gate fidelity or decoherence.}

\vspace{.5cm}
\noindent
{\bf Discussion}

\vspace{.1cm}
\noindent
When fabricating superconducting circuits, parameter variations
unavoidably occur, as in any solid-state system. For the
charge-qubit array, $\nu=0$ can be achieved by having
$\Phi_q=\frac{1}{2}\Phi_0$, irrespective of the parameter
variations. Also, $\mu$ can be tuned, via the gate voltage $V_g$, to
the required value, even if $E_{\rm ch}$ varies for different
qubits. For varying $E_{J0}$ among qubits, the interqubit couplings
also vary [see equation~(\ref{T-1})]. One can replace the large
Josephson junction by a dc SQUID and tune the SQUID, i.e., the
effective $E_{Jc}$, to obtain the desired value $t$ for the
interqubit coupling. For a symmetric SQUID with Josephson coupling
energy $E_{Jc}^{(0)}$, the effective $E_{Jc}$ is given by
$E_{Jc}(\Phi_c)=2E_{Jc}^{(0)}\cos(\pi\Phi_c/\Phi_0)$, where $\Phi_c$
is the magnetic flux in the SQUID loop. In equation~(\ref{T-1}),
$E_{Jc}$ is now replaced by $E_{Jc}(\Phi_c)$; in both
equation~(\ref{T-1}) and $\nu$, $\Phi_q$ is replaced by
$\Phi'_q\equiv\Phi_q+\frac{1}{2}\Phi_c$. Therefore, the tunability
of $t(\Phi_c,\Phi'_q)$ with $\nu=0$ can be implemented by
changing both $\Phi_c$ and $\Phi_q$.

As for the flux-qubit array, $\nu=0$ can be achieved by
having $f=\frac{1}{2}$. Also, $\mu$ can be tuned to the given value
by changing the flux $f_s$ applied to the SQUID in each qubit.
Moreover, even if the parameters of couplers vary, one can tune the
flux $f_c$ in each coupler to achieve the required value of $t$ for
the interqubit coupling [see equation~(\ref{T-2})].

Furthermore, note that even if $\nu=0$ cannot be
experimentally reached very accurately, our proposal still works for
the states in the unpaired-Majorana region of the system if
$|\nu/t|\ll 1$. Experimentally, this can be achieved by designing a
relatively strong interqubit coupling for the qubit array.
For instance, because $I_p\sim 2\pi E_J/\Phi_0$, one has
$|\nu/t|\sim [2\pi(1+2\beta)E_J/\beta
E_{Jc}]|\frac{1}{2}-f|$. When $\beta=0.1$, $E_{Jc}=5E_J$, and $f\in
[0.499,0.501]$, $|\nu/t|\sim 0.015\ll 1$.

In conclusion, we propose superconducting circuits to construct two
superconducting-qubit arrays where Majorana modes can occur. The
unpaired zero-energy Majorana modes, {which emerge at the left and right ends of
the chain in the Majorana-fermion representation, can be used to encode a qubit called the Majorana qubit.
Also, we express this Majorana qubit in the spin representation and show its advantage, over a single superconducting
qubit, for quantum coherence.} Moreover, we suggest using four superconducting qubits as the
smallest circuit to demonstrate the braiding of Majorana modes, and
show how to distinguish the states before and after braiding
Majorana modes. These superconducting-qubit arrays can, in
principle, be extended to wire networks, similar to the
semiconducting wire networks in Ref.~\cite{Alicea}, to demonstrate the
non-Abelian statistics of Majorana modes.


\vspace{.5cm}
\noindent
{\bf Acknowledgement}

\vspace{.01cm}
\noindent
We thank the KITPC for hospitality during the early stage of
this work. J.Q.Y. is supported by the NSFC Grant No.~91121015,
the MOST 973 Program Grant No.~2014CB921401 and the NSAF Grant No. U1330201.
Z.D.W. is supported by the GRF (HKU7045/13P) and the CRF (HKU8/11G)
of Hong Kong. F.N. is partially supported by the RIKEN iTHES Project, MURI Center
for Dynamic Magneto-Optics, and a Grant-in-Aid for Scientific Research (S).


\begin{thebibliography}{99}
%
%
\bibitem{Wilczek} Wilczek, F. Majorana returns. {\it Nat. Phys.} {\bf 5} 614-618 (2009).
\bibitem{Stern} Stern, A. Non-Abelian states of matter. {\it Nature} {\bf 464} 187-193 (2010).
\bibitem{Read} Read, N. \& Green, D. Paired states of fermions in two dimensions with breaking of parity and time-reversal symmetries and the fractional quantum Hall effect. {\it Phys. Rev.} B {\bf 61} 10267-10297 (2000).
\bibitem{Ivanov} Ivanov, D. A. Non-Abelian statistics of half-quantum vortices in p-wave superconductors. {\it Phys. Rev. Lett.} {\bf 86} 268-271 (2001).
\bibitem{Kitaev-1} Kitaev, A. Yu. Unpaired Majorana fermions in quantum wires. {\it Phys. Usp} {\bf 44} 131-136 (2001).
\bibitem{Fu} Fu, L. \& Kane, C. L. Superconducting proximity effect and Majorana fermions at the surface of a topological insulator. {\it Phys. Rev. Lett.} {\bf 100} 096407 (2008).
\bibitem{Sato} Sato, M., Takahashi, Y. \& Fujimoto, S. Non-Abelian topological order in s-wave superfluids of ultracold fermionic atoms. {\it Phys. Rev. Lett.} {\bf 103} 020401 (2009).
\bibitem{Sau} Sau, J. D., Lutchyn, R. M., Tewari, S. \& Das Sarma, S. Generic new platform for topological quantum computation using semiconductor heterostructures. {\it Phys. Rev. Lett.}
{\bf 104} 040502 (2010).
\bibitem{Alicea-PRB} Alicea, J. Majorana fermions in a tunable semiconductor device. {\it Phys. Rev.} B {\bf 81} 125318 (2010).
\bibitem{Lee} Potter, A. C. \& Lee, P. A. Multichannel generalization of Kitaev's Majorana end states and a practical route to realize them in thin films. {\it Phys. Rev. Lett.} {\bf
105} 227003 (2010).
\bibitem{Rakhmanov} Rakhmanov, A. L., Rozhkov, A. V. \& Nori, F. Majorana fermions in pinned vortices.
{\it Phys. Rev.} B {\bf 84} 075141 (2011).
\bibitem{wire1} Lutchyn, R. M., Sau, J. D. \& Das Sarma, S. Majorana fermions and a topological phase transition in semiconductor-superconductor heterostructures. {\it Phys. Rev. Lett.} {\bf 105} 077001 (2010);
\bibitem{wire2} Oreg, Y., Refael, G. \& von
Oppen, F. Helical liquids and Majorana bound states in quantum wires. {\it Phys. Rev. Lett.} {\bf 105} 177002 (2010); \bibitem{wire3} Lutchyn, R. M., Stanescu, T. D.
\& Das Sarma, S. Search for Majorana fermions in multiband semiconducting nanowires. {\it Phys. Rev. Lett.} {\bf 106} 127001 (2011).
{
\bibitem{Mourik} Mourik, V. {\it et al.}
Signatures of Majorana fermions in hybrid superconductor-semiconductor nanowire devices. {\it Science} {\bf 336}, 1003-1007 (2012).
}
\bibitem{Alicea} Alicea, J., Oreg, Y., Refael, G., von Oppen, F. \& Fisher, M. P. A. Non-Abelian statistics and topological quantum information processing in 1D wire networks. {\it Nat. Phys.}
{\bf 7} 412-417 (2011).
\bibitem{Kitaev-2} Kitaev, A. \& Laumann, C. Topological phases and quantum computation. arXiv:~0904.2771
\bibitem{Lieb} Lieb, E., Schultz, T. \& Mattis, D. Two soluble models of an antiferromagnetic chain. {\it Ann. Phys.} (N.Y.) {\bf 16} 407-466 (1961).
\bibitem{You-Nature} You, J. Q. \& Nori, F. Atomic physics and quantum optics using
superconducting circuits. {\it Nature} {\bf 474} 589-597 (2011);
\bibitem{You-PT} You, J. Q. \& Nori, F. Superconducting circuits and
quantum information. {\it Phys. Today} {\bf 58}~(No. 11) 42-47 (2005);
\bibitem{Clarke} Clarke, J. \& Wilhelm, F. K. Superconducting quantum bits. {\it Nature} {\bf 453} 1031-1042 (2008).
{
\bibitem{D-wave} van der Ploeg, S. H. W. {\it et al.}
Controllable coupling of superconducting flux qubits. {\it Phys. Rev. Lett.} {\bf 98}, 057004 (2007).
\bibitem{Hime} Hime, H. {\it et al.}
Solid-state qubits with current-controlled coupling. {\it Science} {\bf 314}, 1427-1429 (2006).
\bibitem{Tsai} Niskanen, A. O. {\it et al.}
Quantum coherent tunable coupling of superconducting qubits. {\it Science} {\bf 316}, 723-726 (2007).
}
\bibitem{Buluta}
Georgescu, I. M., Ashhab, S., \& Nori, F. Quantum Simulation.
{\it Rev. Mod. Phys.} {\bf 86}, 153-186 (2014).
\bibitem{YTN-PRB} You, J. Q., Tsai, J. S. \& Nori, F. Controllable manipulation and entanglement of macroscopic quantum states in coupled charge qubits. {\it Phys. Rev.} B {\bf 68} 024510 (2003).
\bibitem{Grajcar} Grajcar, M., Liu, Y. X., Nori, F. \& Zagoskin, A. M. Switchable resonant coupling of flux qubits. {\it Phys. Rev.} B {\bf 74} 172505 (2006).
\bibitem{micromaser} You, J. Q., Liu, Y. X., Sun, C. P. \& Nori, F. Persistent single-photon production by tunable on-chip micromaser with a superconducting quantum circuit. {\it Phys. Rev.} B {\bf 75} 104516 (2007).
\bibitem{Greenberg} Greenberg, Y. S. {\it et al.}
Low-frequency characterization of quantum tunneling in flux qubits. {\it Phys. Rev.} B {\bf 66} 214525 (2002).
\bibitem{Levitov} Levitov, L. S., Orlando, T. P., Majer, J. B. \& Mooij, J. E. Quantum spin chains and Majorana states in arrays of coupled qubits. arXiv:~cond-mat/0108266.
\bibitem{NJP} DeGottardi, W., Sen, D. \& Vishveshwara, S. Topological phases, Majorana modes and quench dynamics in a spin ladder system. {\it New J. Phys.} {\bf 13} 065028 (2011).
\bibitem{deco1} Budich, J. C., Walter, S. \& Trauzettel, B. Failure of protection of Majorana based qubits against decoherence. {\it Phys. Rev.} B {\bf 85} 121405 (2012).
\bibitem{deco2} Gangadharaiah, S., Braunecker, B., Simon, P. \& Loss, D. Majorana edge states in interacting one-dimensional systems. {\it Phys. Rev. Lett.} {\bf 107} 036801 (2011);
\bibitem{deco3} Stoudenmire, E. M., Alicea, J., Starykh, O. A. \& Fisher,
M. P. A. Interaction effects in topological superconducting wires supporting Majorana fermions. {\it Phys. Rev.} B {\bf 84} 014503 (2011);
\bibitem{deco4} Sela, E., Altland, A. \& Rosch, A. Majorana fermions in strongly interacting helical liquids. {\it Phys. Rev.} B {\bf 84} 085114 (2011).
\bibitem{Tserk} Tserkovnyak, Y. \& Loss, D. Universal quantum computation with ordered spin-chain networks. {\it Phys. Rev.} A {\bf 84}, 032333 (2011).
{
\bibitem{Palmer} Zaretskey, V., Suri, B., Novikov, S., Wellstood, F. C. \& Palmer, B. S.
Spectroscopy of a Cooper-pair box coupled to a two-level system via charge and critical current.
{\it Phys. Rev.} B {\bf 87}, 174522 (2013).
}
{
\bibitem{Khvesh} Khveshchenko, D. V. Entanglement and decoherence in near-critical qubit chains.
{\it Phys. Rev.} B {\bf 68}, 193307 (2003).
}
{
\bibitem{Pedrocchi} Pedrocchi, F. L., Chesi, S., Gangadharaiah, S. \& Loss,
D. Majorana states in inhomogeneous spin ladders. {\it Phys. Rev.} B {\bf 86}, 205412 (2012).
}
{
\bibitem{Kitaev-3} Kitaev, A. Anyons in an exactly solved model and beyond. {\it Ann. Phys.} {\bf 321}, 2-111 (2006).
}
\bibitem{transmon} Koch, J. {\it et al.}
Charge-insensitive qubit design derived from the Cooper pair box. {\it Phys. Rev.} A {\bf 76} 042319 (2007).
\bibitem{low-decoherence} You, J. Q., Hu, X., Ashhab, A. \& Nori, F. Low-decoherence flux qubit. {\it Phys. Rev.} B {\bf 75} 140515 (2007).
\bibitem{enhanced-qubit} Steffen, M. {\it et al.}
High-coherence hybrid superconducting qubit. {\it Phys. Rev. Lett.} {\bf 105} 100502 (2010).
\bibitem{Steffen} Steffen, M. {\it et al.}
Measurement of the Entanglement of Two Superconducting Qubits via State Tomography. {\it Science} {\bf 313} 1423-1425 (2006).
\bibitem{Filipp} Filipp, S. {\it et al.}
Two-qubit state tomography using a joint dispersive readout. {\it Phys. Rev. Lett.} {\bf 102} 200402 (2009).
{
\bibitem{DiCarlo} DiCarlo, L. {\it et al.}
Preparation and measurement of three-qubit entanglement in a superconducting circuit. {\it Nature} {\bf 467}, 574-578 (2010).
\bibitem{Barends} Barends, R. {\it et al.}
Superconducting quantum circuits at the surface code threshold for fault tolerance. {\it Nature} {\bf 508}, 500-503 (2014).
}

\end{thebibliography}
\end{document}